\documentclass{IEEEtran}
\usepackage{cite}
\usepackage{amsmath,amssymb,amsfonts}
\usepackage{graphicx}
\usepackage{textcomp,nicefrac}
\def\BibTeX{{\rm B\kern-.05em{\sc i\kern-.025em b}\kern-.08em
T\kern-.1667em\lower.7ex\hbox{E}\kern-.125emX}}
\usepackage[colorlinks,linkcolor=blue,anchorcolor=blue,citecolor=blue]{hyperref}
\usepackage[switch]{lineno}

\usepackage[LGRgreek]{mathastext}
\MTgreekfont{cmr}
\mathastext[upright]

\begin{document}

\title{Ultra-Fast 4H-SiC AC-LGAD for 4D Tracking Applications}
\author{Tao Yang, Chengxi Yang, Ben J. Sekely, Yashas Satapathy, Abraham Tishelman-Charny, Greg Allion, Gil Atar, Philip Barletta, Gabriele Giacomini, Carl Haber, Steve Holland, John F. Muth, Spyridon Pavlidis, Haichen Wang and Stefania Stucci
\thanks{This work is supported by the U.S. Department of Energy, Office of Science, Office of High Energy Physics, under Contract Number DE-AC02-05CH11231 and Award Number DE-SC0024252. Additional support was provided by the National Science Foundation under contract No. ECCS-1542015.}
\thanks{Tao Yang, Steve Holland and Carl Haber (e-mail: chhaber@lbl.gov) are with Physics Division, Lawrence Berkeley National Laboratory, 1 Cyclotron Road, Berkeley, 94720, CA, U.S.}
\thanks{Ben J. Sekely, Yashas Satapathy, Greg Allion, Gil Atar, Philip Barletta, Spyridon Pavlidis and John F. Muth (e-mail: muth@ncsu.edu) are with Department of Electrical and Computer Engineering, North Carolina State University, 2410 Campus Shore Dr, Raleigh, 27606, NC, U.S. The current affiliation of Ben J. Sekely is NRL/National Research Council, Washington, DC 20375, U.S.}
\thanks{Chengxi Yang and Haichen Wang are with Physics Department, University of California, Berkeley, 94720, CA, U.S.}
\thanks{Abraham Tishelman-Charny, Gabriele Giacomini and Stefania Stucci are with Brookhaven National Laboratory, 98 Rochester St, Upton, 11973, NY, U.S.}
}

\maketitle

\begin{abstract}
4H-SiC low gain avalanche detectors (LGADs) combine internal gain with the high carrier saturation drift velocity, thermal stability, and low leakage current of SiC. AC-coupled readout (AC-LGAD) adds position information to the precise timing, which enables 4D tracking. This work reports the first experimental demonstration of a 4H-SiC AC-LGAD. Devices with a $1.0\times1.0$~mm$^2$ active area and a $2\times2$ array of AC-coupled pads at a pitch of 500~$\mu$m were fabricated with two gain-layer doping concentrations, $4\times10^{17}$ and $5\times10^{17}$~cm$^{-3}$, and characterized with a 375~nm ultraviolet transient current technique (UV-TCT) scanning setup. The hit position and time were reconstructed with the conventional discretized positioning circuit (DPC) and constant fraction discrimination (CFD) methods. They were also reconstructed with a transformer-based neural network, which processes the full waveforms of the four pads through a shared encoder and formulates timing as a per-sample segmentation task. With DPC and CFD, the best position and time resolutions are about 10~$\mu$m and 20~ps. The neural network mitigates the nonlinear distortions near the device edges and improves the position resolution to below 5~$\mu$m and the time resolution to about 15~ps, improvements of more than 60\% and 30\%, respectively. The higher-doped device reaches a high signal-to-noise ratio at lower bias and keeps good performance at low laser intensities. These results demonstrate the potential of 4H-SiC AC-LGADs for 4D tracking in future collider experiments.
\end{abstract}


\section{Introduction} \label{sec:introduction}
Silicon-based low gain avalanche detectors (LGADs) have been extensively studied over the past decade due to their excellent timing performance \cite{CNM_LGAD,HPK_LGAD,FBK_LGAD,BNL_LGAD,KeweiWU_LGAD,YunyunFAN_NDL}. However, conventional Si LGADs suffer from a significant dead zone caused by the junction termination extension (JTE) \cite{JTE_Sim}. To address this issue, AC-LGADs \cite{AC_LGAD} and TI-LGADs \cite{TI_LGAD} were proposed and demonstrated to enable 4D tracking with both time and position resolution \cite{Beam_AC_LGAD,Comprehensive_TI_LGAD}. Compared to silicon, 4H-SiC offers advantages for collider detectors due to its higher carrier saturation drift velocity, superior high-temperature stability, and lower leakage current after irradiation, making it an attractive alternative. Based on these advancements, 4H-SiC LGADs have been proposed and successfully fabricated by several groups \cite{Tao_EDL,SVIHRA_SiC_LGAD_2025,Zhao_SiC_LGAD_2025}, and excellent timing performance has been demonstrated \cite{Tao_EDL}. Recently, DC-coupled segmented 4H-SiC detectors have demonstrated good position resolution and linearity \cite{SiC_Pos}. However, compared with DC coupling, AC coupling provides isolation between the detector high-voltage bias and the front-end electronics. This configuration also blocks the DC leakage current from flowing into the readout amplifier, preventing baseline shifts and mitigating the associated shot-noise contribution. It also relaxes the requirements on the bias current handling, which is advantageous for low-noise and high-precision timing measurements. Furthermore, the internal signal amplification of LGADs enhances the signal-to-noise ratio, making 4H-SiC AC-LGADs a promising candidate for 4D tracking applications. Recent simulation studies based on Weightfield2 (WF2) further suggest the promising radiation tolerance of 4H-SiC LGADs in high-fluence environments \cite{Sim_SiC_AC_LGAD}.

Motivated by the potential of 4H-SiC LGADs for 4D tracking, this work reports the first experimental demonstration of a 4H-SiC AC-LGAD. A 4H-SiC AC-LGAD with $2\times2$ AC-coupled pads is designed and fabricated based on the 4H-SiC LGAD. The position recognition and timing performance of the 4H-SiC AC-LGAD are characterized using a UV-TCT system to demonstrate its excellent 4D tracking capability.

\section{Device Structure} \label{sec:device_structure}

\begin{figure}[!t]
\centerline{\includegraphics[width=0.95\columnwidth]{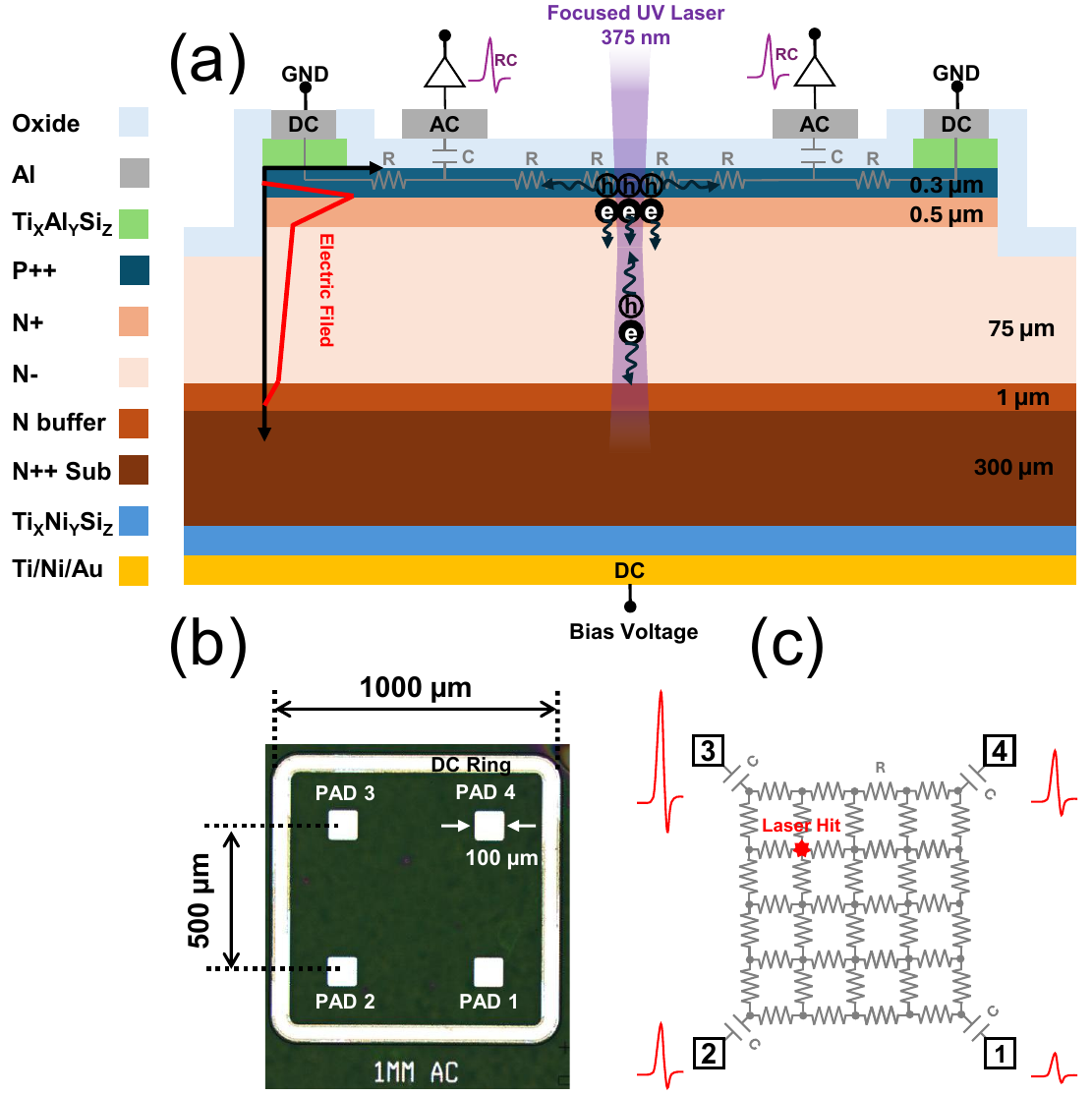}}
\caption{Schematic illustration of the 4H-SiC AC-LGAD. (a) Cross-sectional view of the device structure and readout scheme under focused 375 nm UV laser injection. The laser-induced carriers drift in the depleted region, and the transient signal is capacitively coupled to the AC pads through the resistive layer. (b) Optical image of the fabricated device. (c) Equivalent RC network of the AC-coupled readout, where the induced signal is shared among the four pads according to the laser hit position.}
\label{fig:devide_structure}
\end{figure}

As shown in Fig.~\ref{fig:devide_structure}(a), the epitaxial layer structure of the 4H-SiC AC-LGADs in this work is identical to that reported in \cite{Tao_EDL}. The devices are fabricated on custom-designed 4H-SiC epitaxial wafers. The N$^-$ drift layer has a doping concentration of $2 \times 10^{14}~\text{cm}^{-3}$, and the P$^{++}$ layer, with a doping concentration of $1 \times 10^{19}~\text{cm}^{-3}$, provides ohmic contact to the Ti/Al/Si alloy. Two AC-LGAD devices with different N$^+$ gain-layer doping concentrations, $4 \times 10^{17}~\text{cm}^{-3}$ and $5 \times 10^{17}~\text{cm}^{-3}$, are investigated in this work. Fig.~\ref{fig:devide_structure}(b) shows a photograph of the 4H-SiC AC-LGAD. The active area is $1.0~\text{mm} \times 1.0~\text{mm}$, with four AC pads arranged at a pitch of $500~\mu\text{m}$, and each square pad has a size of $100~\mu\text{m} \times 100~\mu\text{m}$. The outer DC ring contacts the P$^{++}$ layer and, together with the back electrode, reverse-biases the device. The discrete metal pads serve as AC-coupled electrodes through a 300~nm oxide layer. Under the applied bias, the electric field in the gain layer reaches approximately 4~MV/cm, enabling avalanche multiplication of electrons and holes. For a comparative study, we also prepare a 4H-SiC PIN device with the same structure, doping concentrations, and fabrication process, except that it has no gain layer.

\begin{figure}[!t]
\centerline{\includegraphics[width=0.95\columnwidth]{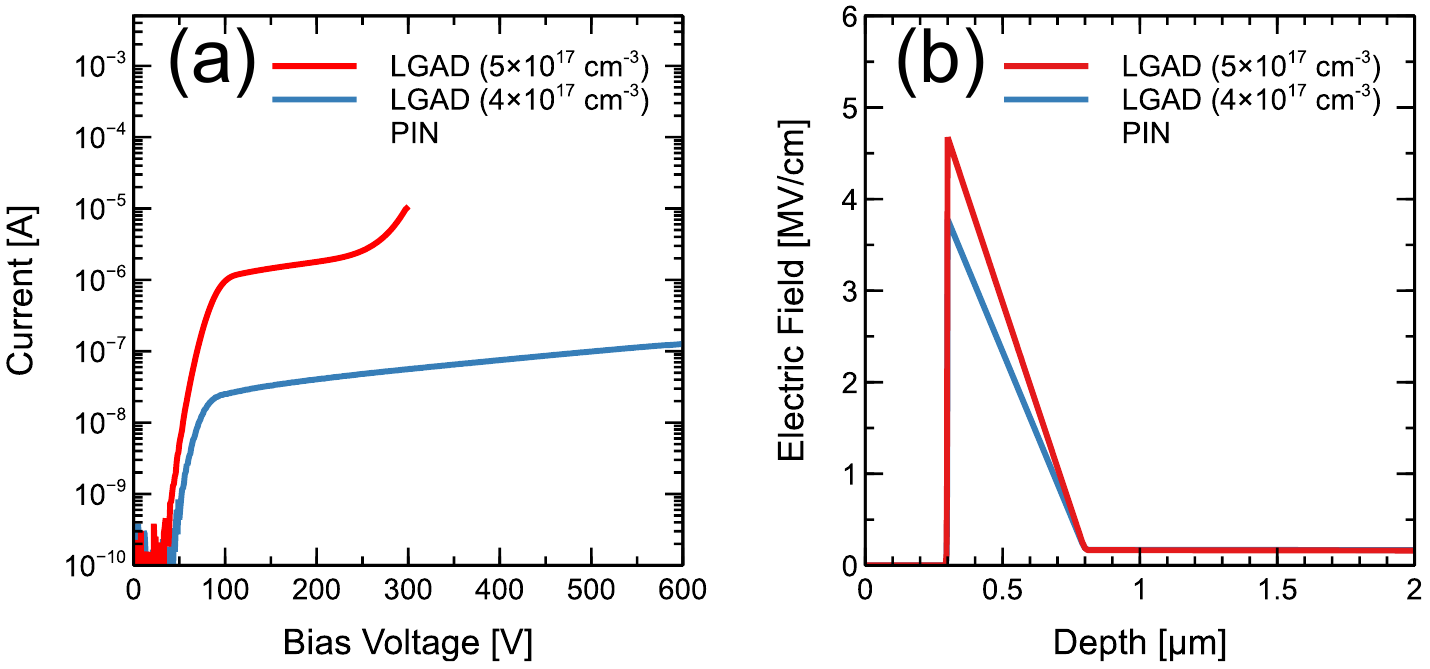}}
\caption{Electrical characterization and simulated electric-field profiles of the 4H-SiC AC-LGADs and the reference PIN device. (a) Measured current-voltage characteristics of the AC-LGADs with gain-layer doping concentrations of $4 \times 10^{17}~\text{cm}^{-3}$ and $5 \times 10^{17}~\text{cm}^{-3}$, together with the PIN device. (b) Simulated electric-field distributions as a function of depth under 500~V reverse bias.}
\label{fig:iv}
\end{figure}

Fig.~\ref{fig:iv} compares the current-voltage characteristics and simulated electric-field distributions of the two 4H-SiC AC-LGADs and the reference PIN device. The low-doping and high-doping LGADs correspond to gain-layer doping concentrations of $4 \times 10^{17}~\text{cm}^{-3}$ and $5 \times 10^{17}~\text{cm}^{-3}$, respectively. As shown in Fig.~\ref{fig:iv}(a), the PIN device keeps a low leakage current over the measured bias range, while the LGAD devices show a clear increase in current as the reverse bias increases. This behavior is consistent with the presence of the gain layer, where the enhanced electric field promotes impact ionization and carrier multiplication. Fig.~\ref{fig:iv}(b) shows that the electric field concentrates in the gain layer for both LGADs. The high-doping LGAD reaches a larger peak electric field than the low-doping LGAD, while the PIN device does not form a comparable high-field multiplication region.

Signal generation in 4H-SiC AC-LGADs starts from the prompt creation of electron-hole pairs by UV laser within the detector (see Fig.~\ref{fig:devide_structure}). The gain layer sustains a high electric field, triggering impact ionization and providing intrinsic current gain of the generated charge. The amplified current induces signals on the surface AC pads through capacitive coupling, while the DC bias is isolated by the AC coupling network. Due to the finite lateral impedance and the $2\times2$ arrangement of square pads, the induced signal is distributed among multiple pads (charge sharing), enabling position localization and, together with the fast timing from the gain mechanism, 4D (time and position) readout. The total collected charge is proportional to the primary ionization and the gain, with the readout electronics sensing mainly the AC component. On each AC pad, the avalanche-generated charge rapidly charges the pad node via capacitive coupling and is subsequently discharged through the AC coupling capacitor and the input impedance of the readout amplifier, with the pulse shape governed by the RC time constants of the pad network and front-end.

\section{Test Setup} \label{sec:test_setup}

\begin{figure*}[!t]
\centerline{\includegraphics[width=2\columnwidth]{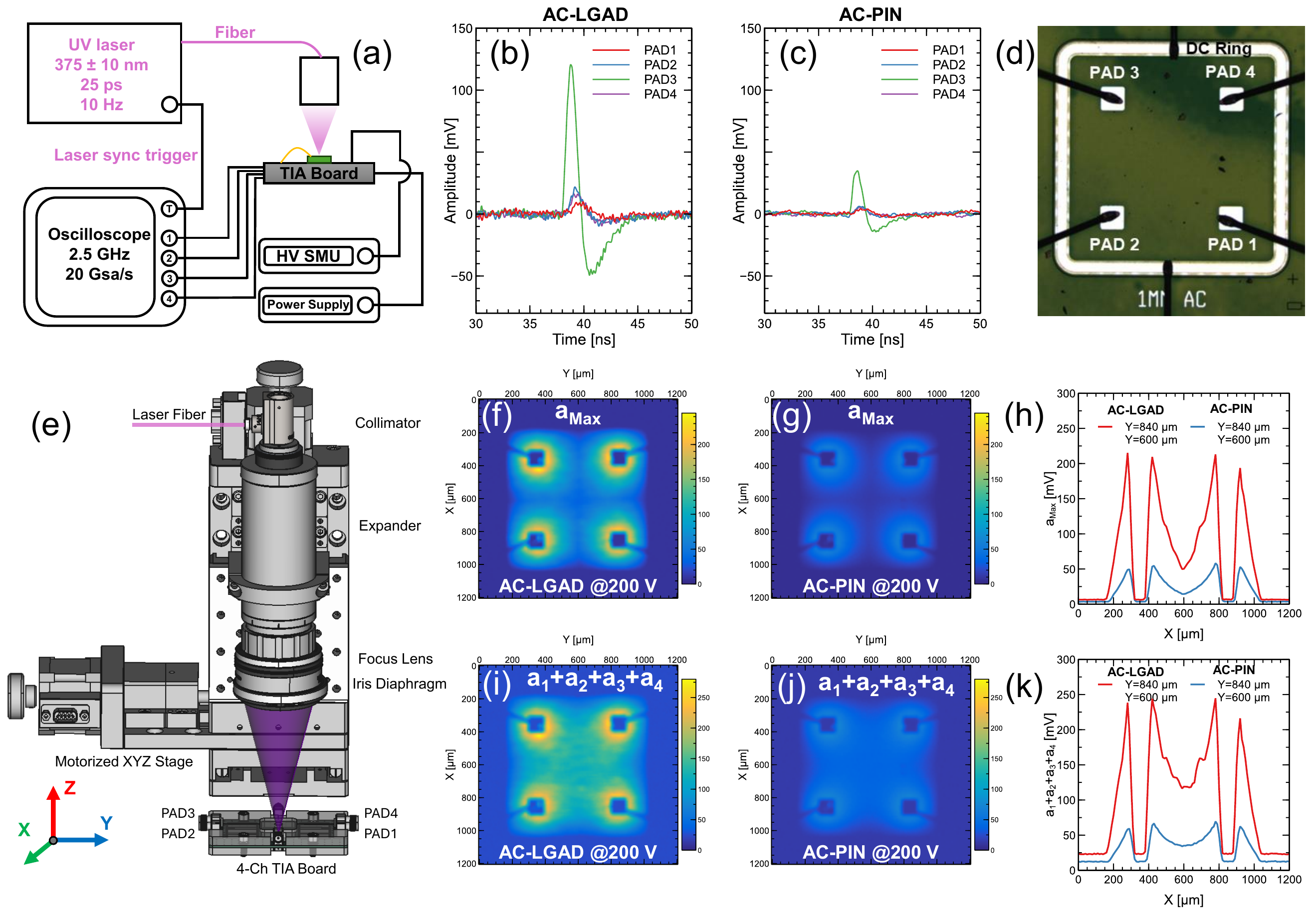}}
\caption{UV-TCT characterization of the 4H-SiC AC-LGAD and the reference 4H-SiC AC-PIN device. (a) Schematic of the UV-TCT setup based on a $375 \pm 10$~nm, 25~ps pulsed laser and a four-channel readout chain. (b,c) Representative laser-induced transient signals measured from the four AC pads of the AC-LGAD and AC-PIN devices, respectively. (d) Optical image of the device under test, showing the four AC pads and the surrounding DC ring. (e) Mechanical configuration of the focused laser scanning system and the four-channel TIA readout board. (f,g) Two-dimensional distributions of the maximum signal amplitude, $a_{\text{Max}}$, measured at 200~V for the AC-LGAD and AC-PIN devices, respectively. (h) Line profiles of $a_{\text{Max}}$ at representative $Y$ positions. (i,j) Two-dimensional distributions of the summed amplitudes, $a_1+a_2+a_3+a_4$, measured at 200~V for the AC-LGAD and AC-PIN devices, respectively. (k) Line profiles of $a_1+a_2+a_3+a_4$ at representative $Y$ positions.}
\label{fig:test_setup}
\end{figure*}

Fig.~\ref{fig:test_setup}(a) and Fig.~\ref{fig:test_setup}(e) show a UV-TCT system based on a 375~nm, 25~ps-wide ultraviolet pulsed laser to characterize the position and time resolutions of the 4H-SiC AC-LGAD. The laser pulse is coupled into a focusing system via an optical fiber, then collimated, expanded, and focused to form a laser spot with a radius of $\sigma\sim$24~$\mu$m. The 4H-SiC AC-LGAD is mounted on a readout board equipped with a two-stage front-end low-noise linear voltage amplifier with a bandwidth of 2~GHz. The board provides four channels, each wire-bonded to one of the device's four AC pads, shown in Fig.~\ref{fig:test_setup}(d). When a laser pulse is injected into the 4H-SiC LGAD, it induces current signals on the four pads and the amplitude of the pulsed current depends on the laser injection position.

As shown in Fig.~\ref{fig:test_setup}(b) and (c), the injected laser pulse induces current pulses in the LGAD and PIN devices. Fig.~\ref{fig:test_setup}(f), (g), and (h) show the distributions of the maximum signal values from the four pads, which reflect the correlation between the signal pulse amplitude and position. Fig.~\ref{fig:test_setup}(i), (j), and (k) show the distributions of the sum of the signal amplitudes from the four pads, which reflect the correlation between the total signal charge and position. The above results show that the signal does not have an ideal linear dependence on position. The total signal charge is also not uniform, since part of the charge leaks through the external DC ring. These effects make the reconstruction of timing and position information more challenging. Therefore, in this work, we combine conventional numerical analysis with machine learning to reconstruct the timing and position information.

\section{Methodology} \label{sec:methodology}

\subsection{Discretized Positioning Circuit (DPC)}
According to DPC, a dataset of signals with known position information allows us to establish the relationship between position and signal amplitude based on Eq.~\eqref{eq:dpc}:

\begin{equation} \label{eq:dpc}
    \begin{cases}
    x_{\text{Rec}} = x_0 + k_x m \\[6pt]
    m = \dfrac{a_1 + a_2 - a_3 - a_4}{a_1 + a_2 + a_3 + a_4} \\[12pt]
    k_x = \ell \dfrac{\sum (m_{i+1} - m_i)}{\sum (m_{i+1} - m_i)^2}
    \end{cases}
    \!
    \begin{cases}
    y_{\text{Rec}} = y_0 + k_y n \\[6pt]
    n = \dfrac{a_1 + a_4 - a_2 - a_3}{a_1 + a_2 + a_3 + a_4} \\[12pt]
    k_y = \ell \dfrac{\sum (n_{i+1} - n_i)}{\sum (n_{i+1} - n_i)^2}
    \end{cases}
\end{equation}

where $x_{\text{Rec}}$ and $y_{\text{Rec}}$ denote the reconstructed positions along the $x$ and $y$ directions, respectively. $x_0$ and $y_0$ are the reference positions, while $k_x$ and $k_y$ are the calibration coefficients obtained from the signal dataset with known positions. The variables $m$ and $n$ are the normalized amplitude differences along the two directions, and $a_1$, $a_2$, $a_3$, and $a_4$ represent the signal amplitudes measured from the four pads. The parameter $\ell$ denotes the scanning step size between two adjacent laser injection positions.

\begin{equation} \label{eq:coordinate_correction}
    \left\{
    \begin{array}{l}
    x'_{\text{Hit}} = x_{\text{Hit}} \cos\theta - y_{\text{Hit}} \sin\theta + \delta x \\
    y'_{\text{Hit}} = x_{\text{Hit}} \sin\theta + y_{\text{Hit}} \cos\theta + \delta y \\
    \end{array}
    \right.    
\end{equation}

To eliminate the offset between the sensor coordinate system and the laser coordinate system, Eq.~\eqref{eq:coordinate_correction} is used to transform the hit positions. The correction parameters $\theta$, $\delta x$, and $\delta y$ are obtained by minimizing $\sum_{\text{points}} \left[ (x'_{\text{Hit}} - x_{\text{Rec}})^2 + (y'_{\text{Hit}} - y_{\text{Rec}})^2 \right]$, where $\theta$ denotes the relative rotation angle between the two coordinate systems, and $\delta x$ and $\delta y$ denote the translational offsets along the $x$ and $y$ directions, respectively. 

Fig.~\ref{fig:posrec}(a) compares the true hit positions with those reconstructed using the DPC method. Significant distortions are observed near the device edges. These nonlinear distortions arise from the combined effects of the electrode geometry and charge leakage to the surrounding DC ring near the edges. Similar behavior has been widely reported in studies of 2$\times$2 AC-LGADs \cite{RSD_TCAD,RSD_4D}. In the central region of the device, these distortions are less pronounced. In the following analysis, shown in Fig.~\ref{fig:posrec}(b), machine learning is employed to mitigate these distortions and improve the position resolution.

\subsection{Constant Fraction Discrimination (CFD)}

The hit time is reconstructed from the timing information measured by the four pads. For the $i$-th pad, the corrected hit time is given by

\begin{equation}
    t_{\text{Hit}}^i = t_{\text{ToA}}^i - t_\text{Offset}^i - d^i/v
\end{equation}

where $t_{\text{ToA}}^i$ is the time of arrival obtained by constant fraction discrimination, $t_\text{Offset}^i$ is the channel-dependent timing offset, which is usually related to differences in cable length among channels. $d^i$ is the distance between the hit position and the $i$-th pad, and $v$ is the effective signal propagation velocity. The term $d^i/v$ corrects the propagation delay from the hit position to the corresponding pad.

The final hit time is calculated as an amplitude-weighted average of the corrected hit times from the four pads,

\begin{equation}
    t_{\text{Hit}} =  \dfrac{\sum_{i=1}^{i=4} t_{\text{Hit}}^i \times a_i}{\sum_{i=1}^{i=4} a_i}
\end{equation}

where $a_i$ denotes the signal amplitude of the $i$-th pad. This weighting gives larger contributions to pads with stronger signals and suppresses the influence of pads with lower signal-to-noise ratios. The time difference with respect to the trigger is then defined as

\begin{equation}
    \Delta t = t_{\text{Hit}} - t_{\text{Trigger}}
\end{equation}

where $t_{\text{Trigger}}$ is the reference trigger time. In this work, the trigger contributes approximately 12~ps of timing jitter \cite{Tao_EDL}.

\subsection{Machine Learning}
\begin{figure*}[!t]
\centering
\includegraphics[width=2\columnwidth]{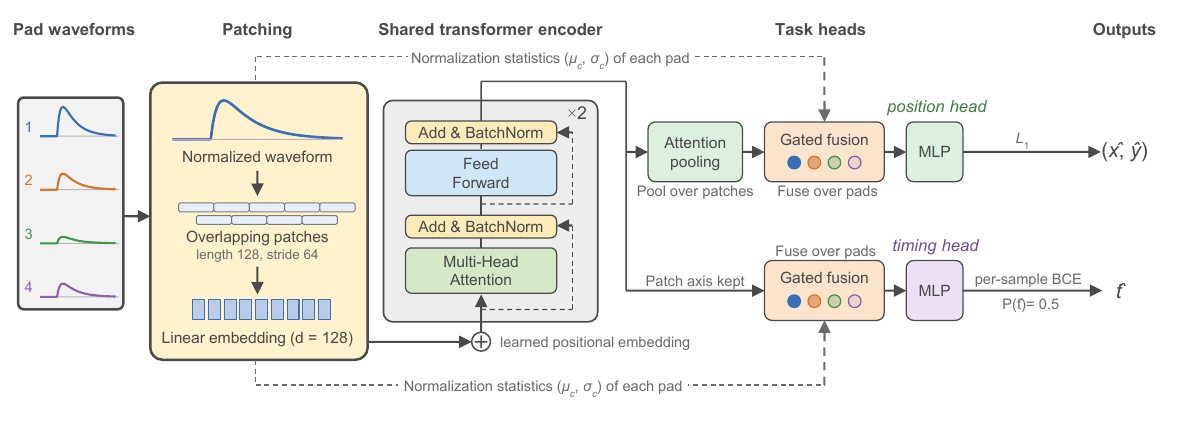}
\caption{Architecture of the ML reconstruction model. A shared transformer encoder processes each pad waveform independently. The per-pad normalization statistics $(\mu_c,\sigma_c)$ enter the gated fusion, which collapses the pad axis. The position head predicts $(x, y)$ and is trained with an $L_1$ loss. The timing head predicts a per-sample probability $P$ of being after the hit; it is trained with a BCE loss, and $\hat{t}$ is the first upward crossing of $P=0.5$.}
\label{fig:ml_model}
\end{figure*}

Compared with the DPC and CFD methods, which use only the peak amplitude and a single threshold crossing of each pad, machine learning can exploit the full waveforms and is therefore expected to improve the reconstruction. We develop a transformer-based neural network that reconstructs both position and time with a shared encoder, which is attractive for future FPGA implementation~\cite{ATLAS_LAr_NN_2021}. As shown in Fig.~\ref{fig:ml_model}, the model consists of a transformer encoder, a gated channel-fusion module, and two task heads for position and time reconstruction. In contrast with previous work on machine learning reconstruction for Si AC-LGADs~\cite{Siviero_MLfirst, Siviero_MLhitpos, Aoki_fullwaveform}, which focused only on position reconstruction, the present model reconstructs both position and time.

The encoder follows the PatchTST architecture~\cite{PatchTST} and is shared by all pads: the waveform from each pad is normalized by its mean and standard deviation~\cite{RevIN}, divided into overlapping patches of 128 samples (0.8~ns) with a stride of 64, linearly embedded, and processed by two transformer blocks with multi-head self-attention~\cite{Attention} and batch normalization. The encoded features, the mean and standard deviation removed by the normalization (which retain the amplitude information needed for charge sharing), and a learned embedding of the pad index are passed to the gated channel-fusion module to form a fused representation.

The two heads take the fused representation as input and are trained jointly with equally weighted losses. The position head pools over patches and regresses $(x,y)$ coordinates with an $L_1$ loss against the laser-stage coordinates. The timing head formulates the time reconstruction as segmentation instead of regression, similar to arrival-time picking in seismology~\cite{PhaseNet}: each waveform is randomly shifted by up to $\pm20$~ns relative to a reference time that is fixed with respect to the laser-driver trigger, and each sample is labeled as lying before (0) or after (1) the reference time. The head predicts this probability for each sample with a binary cross-entropy loss. Finally, the hit time is taken as the point where the probability first crosses 0.5. Compared with scalar regression of hit time \cite{ML_time_discrimination, Wang_MRPC_2020, Ai_2021}, this formulation supervises every sample of the waveform and makes the predicted time follow the pulse in the acquisition window.

A separate model is trained for each sensor, laser intensity, and bias voltage. To obtain an unbiased evaluation of the model performance, the scan grid is divided in a chessboard pattern into interleaved training and validation positions. All waveforms from a validation position are excluded from training, so the reported performance tests interpolation to hit positions unseen in training rather than memorization of repeated waveforms at the same location. This machine learning approach is particularly useful here because the nonuniform charge sharing and position-dependent pulse shapes observed in Fig.~\ref{fig:test_setup} are not fully described by the idealized analytical relations.

\begin{figure}[!t]
\centerline{\includegraphics[width=0.95\columnwidth]{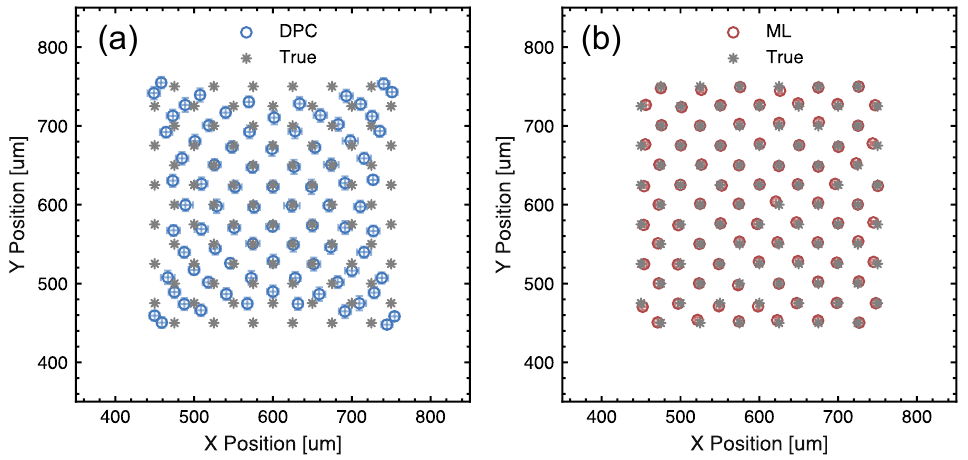}}
\caption{(a) Comparison between the true hit positions and those reconstructed using the analytical DPC method; (b) comparison between the true hit positions and those reconstructed using machine learning.}
\label{fig:posrec}
\end{figure}

\begin{figure*}[!t]
\centerline{\includegraphics[width=2\columnwidth]{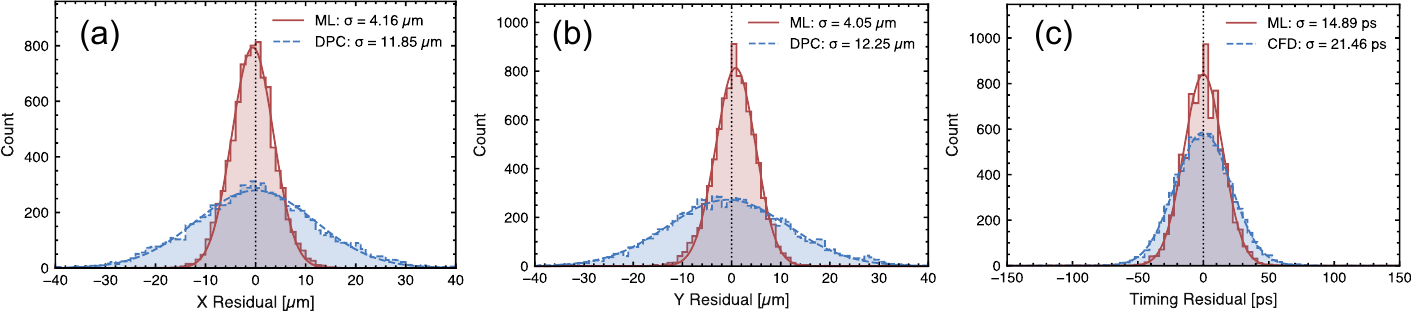}}
\caption{Comparison of the position resolutions obtained using the analytical DPC method and machine learning in (a) the X direction and (b) the Y direction; (c) comparison of the time resolutions obtained using the CFD method and machine learning.}
\label{fig:fit}
\end{figure*}

\begin{figure*}[!t]
\centerline{\includegraphics[width=2\columnwidth]{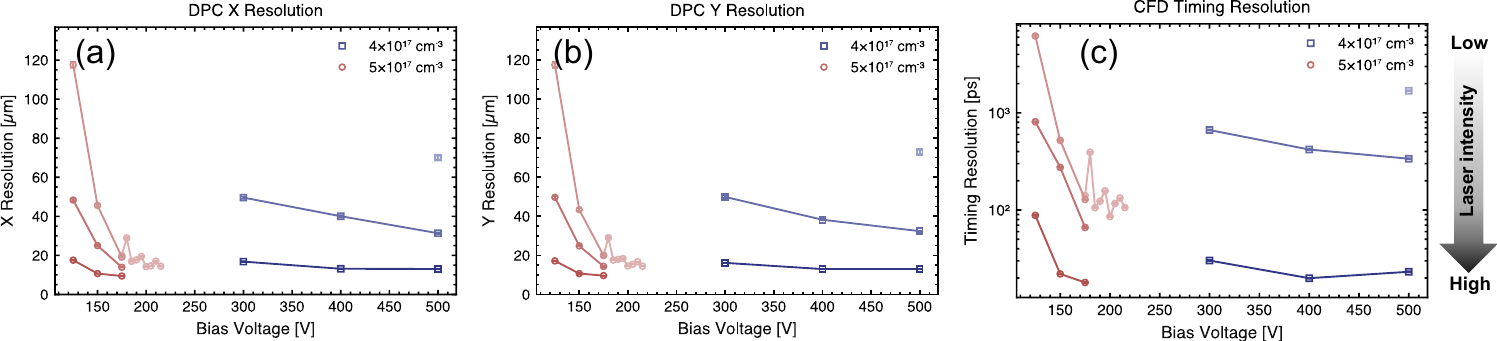}}
\caption{Position and time resolutions as a function of bias voltage at different laser intensities for doping concentrations of $4 \times 10^{17}~\text{cm}^{-3}$ and $5 \times 10^{17}~\text{cm}^{-3}$: (a) X and (b) Y position resolutions obtained using the analytical DPC method; (c) time resolution obtained using the CFD method.}
\label{fig:dpc_cfd}
\end{figure*}

\begin{figure*}[!t]
\centerline{\includegraphics[width=2\columnwidth]{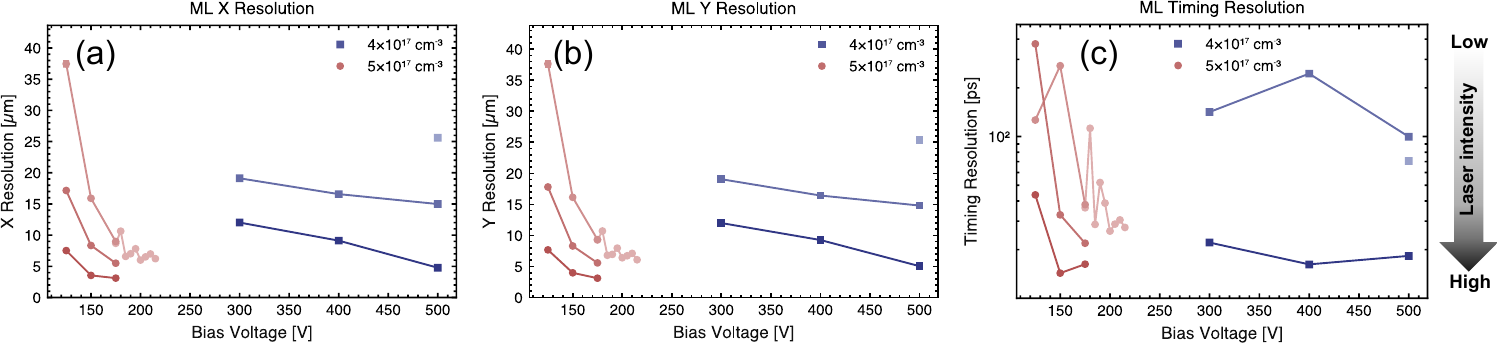}}
\caption{Position and time resolutions obtained using machine learning as a function of bias voltage at different laser intensities for doping concentrations of $4 \times 10^{17}~\text{cm}^{-3}$ and $5 \times 10^{17}~\text{cm}^{-3}$: (a) X position resolution; (b) Y position resolution; and (c) time resolution.}
\label{fig:ml}
\end{figure*}




\section{4D Tracking Performance}

To comprehensively evaluate the device performance, AC-LGADs with doping concentrations of $4 \times 10^{17}~\text{cm}^{-3}$ and $5 \times 10^{17}~\text{cm}^{-3}$ are characterized using both conventional analytical and machine learning reconstruction methods. Fig.~\ref{fig:fit} compares a representative set of experimental results. As shown in Fig.~\ref{fig:fit}(a) and Fig.~\ref{fig:fit}(b), machine learning effectively mitigates distortions near the device edges (see Fig.~\ref{fig:posrec}) and improves the position resolution by more than 60\% compared with the DPC method. Fig.~\ref{fig:fit}(c) shows that machine learning improves the time resolution by more than 30\% compared with the CFD method. The smaller improvement in time resolution reflects the substantial contributions from the signal-to-noise ratio, readout electronics, and reference trigger jitter, which limit the gains achievable through machine learning.

Figs.~\ref{fig:dpc_cfd} and \ref{fig:ml} compare the position and time resolutions of AC-LGADs with two different doping concentrations. For both analytical and machine learning reconstruction methods, the position and time resolutions generally improve with increasing bias voltage and laser intensity. Owing to its higher gain, the device with a doping concentration of $5 \times 10^{17}~\text{cm}^{-3}$ achieves a high signal-to-noise ratio even at relatively low bias voltages. Consequently, its position and time resolutions improve rapidly with increasing bias voltage, and excellent performance is maintained even at low laser intensities. In contrast, the device with the lower doping concentration $4 \times 10^{17}~\text{cm}^{-3}$ exhibits substantially poorer position and time resolutions, particularly at low laser intensities, where the signal-to-noise ratio is low. Under these conditions, a higher bias voltage is required to increase the signal-to-noise ratio.

The measurement results show that, even with conventional analytical reconstruction methods, the 4H-SiC AC-LGAD designed and fabricated in this work achieves a best position resolution approaching 10~$\mu$m and a time resolution of approximately 20~ps. Machine learning further improves the position resolution to below 5~$\mu$m and the time resolution to approximately 15~ps. These results demonstrate the excellent 4D tracking capabilities of 4H-SiC AC-LGADs.

\section{Conclusion}

In this work, 4H-SiC AC-LGADs with a $1.0\times1.0$~mm$^2$ active area and a $2\times2$ array of AC-coupled pads are designed and fabricated, and the first experimental characterization of such devices for 4D tracking is reported. Two devices that differ only in the gain-layer doping concentration, $4\times10^{17}$ and $5\times10^{17}$~cm$^{-3}$, are studied together with a PIN device from the same epitaxial structure, and their position and timing performance is mapped with a 375~nm UV-TCT scanning system over a range of bias voltages and laser intensities.

With the conventional DPC and CFD methods, the best position and time resolutions are about 10~$\mu$m and 20~ps. A transformer-based neural network, which processes the full waveforms of the four pads with a shared encoder and formulates timing as a per-sample segmentation task, mitigates the distortions near the device edges. Evaluated on scan positions excluded from training, it improves the position resolution by more than 60\% to below 5~$\mu$m and the time resolution by more than 30\% to about 15~ps. The smaller improvement in timing reflects the remaining contributions from the signal-to-noise ratio, the readout electronics, and the reference trigger jitter. The comparison between the two devices shows that the gain-layer doping sets the operating window: the higher-doped device reaches a high signal-to-noise ratio at relatively low bias and maintains good resolution at low laser intensities, whereas the lower-doped device requires a higher bias to reach comparable performance.

These results are obtained with UV laser injection and a dedicated model trained for each device and operating condition. Further work includes verification with minimum ionizing particles in beta-source and test-beam measurements, studies of radiation tolerance, optimization of the pad geometry, the resistive layer, and the DC ring to reduce the edge distortions and charge loss, and an FPGA-compatible implementation of the reconstruction model. Overall, the 4H-SiC AC-LGAD demonstrated here is a promising candidate for 4D tracking in future collider experiments.

\newpage
\bibliographystyle{ieeetr}
\bibliography{p5_sic_lgad}

\end{document}